\documentclass[10pt,a4paper,titlepage]{article}
\usepackage[utf8]{inputenc}
\usepackage[english]{babel}
\usepackage{graphicx}
\usepackage{amsmath}
\usepackage{amsfonts}
\usepackage{amsthm}
\usepackage{color}
\usepackage[margin=3.0cm,right=3.0cm,left=3.0cm,twoside]{geometry}

\title{From the elasticity theory to cosmology and vice versa}
\author{Luca Levrino and Angelo Tartaglia}

\begin{document}

\begin{center}
{\LARGE \textbf{From the elasticity theory to cosmology \\ and vice versa}}
\vspace*{0.5cm}

{\large \textit{Luca Levrino}\footnote{E-mail: luca.levrino@asp-poli.it} and \textit{Angelo Tartaglia}\footnote{also at INFN. E-mail: angelo.tartaglia@polito.it}}

\vspace*{0.2cm}
Politecnico di Torino, Department of Applied Science and Technology \\ Corso Duca degli Abruzzi 24 \\ 10129, Torino, Italy \\
\end{center}

\vspace*{2.5cm}

The paper shows how a generalization of the elasticity theory to four dimensions and to space-time allows for a consistent description of the homogeneous and isotropic universe, including the accelerated expansion. The analogy is manifested by the inclusion in the traditional Lagrangian of general relativity of an additional term accounting for the strain induced in the manifold (i.e. in space-time) by the curvature, be it induced by the presence of a texture defect or by a matter/energy distribution. The additional term is sufficient to account for various observed features of the universe and
to give a simple interpretation for the so called dark energy. Then, we show how the same approach can be adopted back in three dimensions to obtain the equilibrium configuration of a given solid subject to strain induced by defects or applied forces. Finally, it is shown how concepts coming from the familiar elasticity theory can inspire new approaches to cosmology and in return how methods appropriated to General Relativity can be applied back to classical problems of elastic deformations in three dimensions.

\vspace*{0.8cm}
deformable solids, cosmology, elastic strain, Lagrangian density and tensors

\vspace*{0.3cm}

\newpage

\section{Introduction}
\label{intro}
In the current perception of the general public (and of most physicists) nothing is as far apart as cosmology, on one side, and everyday life and its related physics, on the other. Of course the basic laws are the same, but the scales are so different that, when considering the universe at large, properties and phenomena emerge which are completely irrelevant at the human scale. It is however true that, when trying to describe and interpret what we see (or our instruments see) in the sky and deep in the past, we necessarily use the tools of reason with which we are equipped as well as the experience we draw from the world around us.
Here we shall show how a phenomenon as familiar as elasticity and its related theory can be useful in the study of cosmology, and in return we shall apply a technique used in general relativity to the solution of simple ordinary problems of elastic deformation. The purpose is to stress the unity of physical sciences and the fruitfulness of the cross-fertilization among different areas of scientific and practical knowledge.
Our start will be the observation that space-time, which is one of the basic ingredients of general relativity, looks like a four-dimensional physical continuum. By physical continuum we mean something endowed with peculiar properties and able to interact with the other ingredient which is matter/energy. The configuration of space-time and the way it reacts to the presence of matter/energy distributions are described in terms of geometrical properties. The same can be said of ordinary material continua in three dimensions. In both cases the best description of the properties and behaviour of our manifolds is cast in the form of tensors. The obvious difference is that for ordinary substances in three dimensions we have a dynamics expressed in terms of an independent parameter, time, which is instead part of the manifold when considering space-time. If we want to maintain a parallelism between material continua and space-time we shall compare `equilibrium' configurations for both cases, i.e. `static' configurations which may be labeled as being of equilibrium, respectively in three and in four dimensions.
Now, how can we distinguish an equilibrium configuration from any other situation? A universally used technique is to write down the action integral appropriated to the system under study. Equivalently, we must consider the Lagrangian density of our manifolds. In fact, the approach we are describing is based on Hamilton's principle, which is a sort of `economy' criterion not directly descending from other basic laws nor emerging from experiments. However, it is elegant and, above all, it works both for classical and quantum fields.
In the case of General Relativity (GR), the Lagrangian density is chosen to be the simplest scalar which can be built from the geometrical tensor expressing the global and local configuration of space-time, i.e. from the Riemann tensor. That quantity is the scalar curvature $R$. In GR the action for an empty space-time is the Einstein-Hilbert action, namely:

\begin{equation}
\label{eq:hilbert}
S=\int{R\sqrt{-g}\,\mbox{d}^4x}
\end{equation}

$\sqrt{-g}\,\mbox{d}^4x$ is the invariant (i.e. independent of the choice of coordinates) volume element in four dimensions.
As for $R$, its meaning is better understood when we remark that for an ordinary bi-dimensional surface it is proportional to the Gaussian curvature, i.e. the inverse product of the two principal curvature radii at a given point.
The Lagrange coordinates of our system, i.e. the generalized coordinates that identify the state of the manifold (space-time), are the components of the metric tensor, i.e. the basic tensor representing the geometric properties of the manifold. The scalar curvature $R$ is a function of the partial derivatives of the elements of the metric tensor up to (linearly) the second order. In a sense we may set a parallelism between the kinetic term in the classical three-dimensional Lagrangian (the kinetic energy), which contains the derivatives of the coordinates with respect to the external parameter (time), and $R$ in four dimensions. However, if we imagine a physical continuum in three-dimensions obtained by deformation from an unstrained state we also have to take into account the elastic deformation energy. By analogy, extending this fact to four dimensions, we are led to adjust the Lagrangian density: to $R$ we sum a scalar function of the Lagrangian coordinates accounting for the strain energy of the manifold. This approach is the essence of the Strained State Cosmology (SSC) theory (see Tartaglia \cite{tartaglia}, and Tartaglia and Radicella \cite{radicella}). The explicit form of the additional energy density $W$ will be worked out in the following. Once the new form of the action integral has been written, we may apply the stationary action principle and see what happens. We find a good description of an accelerated expanding universe: then, we shall verify that the same method works in three dimensions, also reproducing known results usually obtained by other means.

\section{Lagrangian approach to cosmology including the strain of space-time}
\begin{figure}[!ht]
\centering 
\includegraphics[width=129 mm]{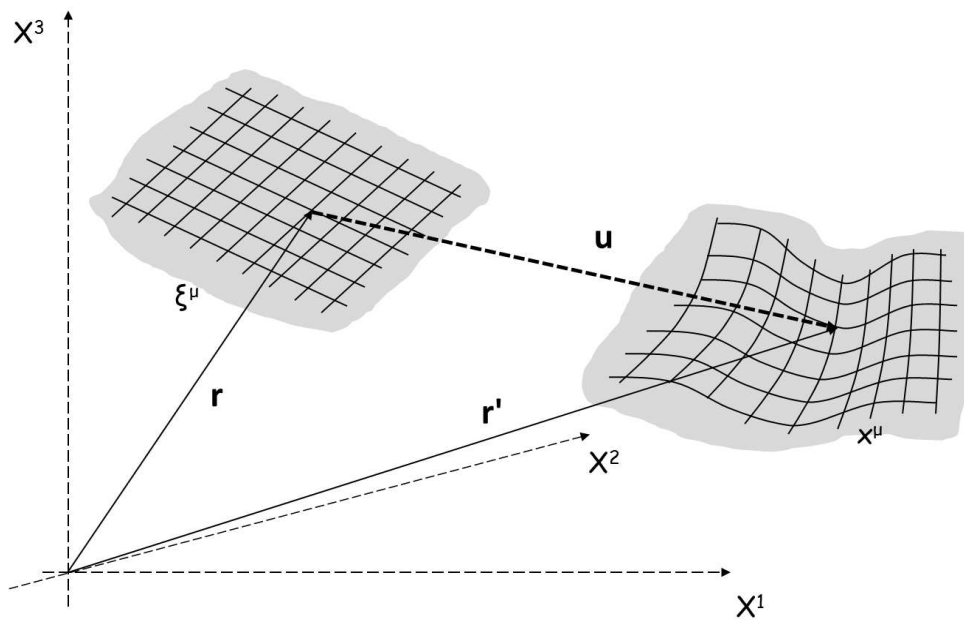}
\caption{Embedding space containing a flat and a curved manifolds}
\label{fig:displ}
\end{figure}

In order to visualize the practical meaning of our approach, we may imagine two separate manifolds: one is a copy of the initial state of the actual manifold of our interest (reference manifold); the other is the natural manifold in its given deformed state. This picture implies that we adopt a higher dimensional point of view. For instance, if the manifolds we are studying are simple bidimensional surfaces, we look at them from within a three-dimensional space: see Fig.\ref{fig:displ}, where also the correspondence between pairs of positions on the two manifolds is shown. The embedding space is flat, which means that for this space Euclidean geometry holds; the same holds true for the reference manifold, but the natural manifold will in general be curved. With more dimensions (say $N$) the scheme is the same and the embedding manifold will be $N+n$ dimensional. If this description is used for space-time, the natural manifold, besides being four-dimensional, will include a Lorentzian signature, i.e. its tangent space at any location will be Minkowskian; putting it in simpler wording, the local speed of light will be the same everywhere, as required by relativity.

We may now consider pairs of corresponding positions arbitrarily close to each other on the two manifolds. Let us compare the corresponding infinitesimal lengths. On the reference manifold we have
\begin{equation}
\label{eq:reference}
\mbox{d} l^2=E_{\mu\nu}\,\mbox{d}\xi^\mu \mbox{d}\xi^\nu
\end{equation}
The choice of the coordinates on the reference is arbitrary, since the result is a true scalar quantity, which means precisely that it does not depend on the coordinates. $E_{\mu\nu}$ are the components of the Euclidean metric tensor in the representation appropriate for the chosen coordinates.

On the natural manifold the metric becomes non-Euclidean\footnote{We shall remark that only the reference manifold is flat, thus the following relation holds only for the metric $E_{\mu\nu}$ ($y$ indicates Cartesian coordinates):
\begin{equation*}
E_{\mu\nu}=\delta_{\alpha \beta} \frac{\partial y^\alpha}{\partial \xi^\mu}\frac{\partial y^\beta}{\partial \xi^\nu}
\end{equation*}

and not for the metric $g_{\mu\nu}$.}:
\begin{equation}
\label{eq:natural}
\mbox{d} l'^2=g_{\mu\nu}\,\mbox{d} x^\mu \mbox{d} x^\nu
\end{equation}
Now, $g_{\mu\nu}$ are the components of the metric tensor on the natural manifold expressed in the coordinates chosen there.

If we suppose that the natural manifold is obtained from the reference one by continuous deformation, we conclude that each position on one manifold has a unique correspondent on the other. If it is so, given an arbitrary choice of coordinates for one of the manifolds, an invertible function will give the corresponding values of the coordinates on the other. In practice, it is possible to express both line elements in terms of just one system of coordinates, for example the one used for the natural manifold.

We can now express the difference $\delta\left( \mbox{d} l^2\right)=\mbox{d} l'^2-\mbox{d} l^2$ between the two line elements (\ref{eq:reference}) and (\ref{eq:natural}) in the same coordinates:

\begin{equation}
\label{eq:differ}
\mbox{d} l'^2-\mbox{d} l^2=(g_{\mu\nu}-E_{\mu\nu})\mbox{d} x^{\mu}\mbox{d} x^{\nu}=2\varepsilon_{\mu\nu}\mbox{d} x^{\mu}\mbox{d} x^{\nu}
\end{equation}

Formula (\ref{eq:differ}) defines the strain tensor $\varepsilon_{\mu\nu}$. It is:

\begin{equation}
\label{eq:metrics}
\varepsilon_{\mu\nu}=\frac{g_{\mu\nu}-E_{\mu\nu}}{2}
\end{equation}

\subsection{Einstein equations including the strain of space-time}
As we have already written, the intrinsic distortion of the natural manifold makes us wonder about the existence of some sort of deformation energy, which must be part of the Lagrangian as an additional potential (and in the equations it has to lead to a dynamical history of the universe as well). Starting from the \textit{ansatz} that our space-time is homogeneous and isotropic\footnote{One remark about homogeneity and isotropy of space-time can be made. In fact, the unstrained manifold is obviously isotropic, since neither exogenous actions nor intrinsic defects are present. When we pass to the natural strained manifold, the locally anisotropic strain can induce some anisotropy in the elastic parameters. Nonetheless, this anisotropy may be considered as of second order with respect to the strain, thus the theory we consider is linear and the conditions of homogeneity and isotropy hold.}, and assuming it behaves linearly, we may apply Hooke's law (whatever the number of dimensions is) in the form:
\begin{equation}
\label{eq:Hooke}
\sigma_{\mu\nu}=\lambda\eta_{\mu\nu}\varepsilon+2\mu\varepsilon_{\mu\nu}
\end{equation}
where $\lambda$ and $\mu$ are the two Lam\'e coefficients, accounting for the properties of the continuum, and $\sigma_{\mu\nu}$ are the components of the stress tensor; $\eta_{\mu\nu}$ are the components of the metric tensor of the flat space-time, i.e. the Minkowski space-time, which is the tangent manifold at any given event of the actual `natural' manifold.

The deformation energy density can then be written as
\begin{equation}
\label{eq:def}
W=\frac{1}{2}\sigma_{\mu \nu}\varepsilon^{\mu \nu}
\end{equation}
Casting (\ref{eq:Hooke}) into (\ref{eq:def}) the final form for the deformation energy is
\begin{equation}
\label{eq:deform en}
W= \frac{1}{2} \lambda \varepsilon^2 + \mu \varepsilon_{\alpha \beta} \varepsilon^{\alpha \beta}
\end{equation}

In practice, we treat space-time  as a physical continuum endowed with properties analogous to the ones of ordinary elastic solids. As a first step, we reconsider the Einstein-Hilbert Lagrangian density, $\mathcal{L}=R \sqrt{-g}$, to which the deformation energy density must be added. Hence,  the complete Lagrangian density \textit{in vacuo} is
\begin{equation}
\label{eq:lag dens SSC}
\mathcal{L}=\left(  R + W \right) \sqrt{-g}
\end{equation}

In general $R \neq 0$, because we assume the presence, somewhere in the manifold, of a defect that generates intrinsic distortion.
From (\ref{eq:lag dens SSC}) new generalized Einstein equations can be written differentiating with respect to the $g_{\mu\nu}$ which are the Lagrangian coordinates of the configuration of space-time. The deformation energy $W$ contributes to a new additional stress-energy tensor, ${T_e}_{\mu\nu}$, which has to be read as an effective elastic stress-energy tensor. Hence, the final equations, the Einstein equations including the strain of space-time, look like
\begin{equation}
\label{eq:Einstein SSC}
G_{\mu\nu} = \kappa\,  T_{\mu\nu} +{T_e}_{\mu\nu}
\end{equation}
Because ${T_e}_{\mu\nu}$ is obtained varying a scalar with respect to a true tensor (the metric tensor $g_{\mu\nu}$), it is also a good tensor, with all the properties of tensors. In explicit form:
\begin{equation}
\label{eq:elastic stress energy}
{T_e}_{\mu\nu}=\lambda \varepsilon \varepsilon_{\mu\nu}+2 \mu \varepsilon_{\mu\nu}
\end{equation}
This equation comes from formula (\ref{eq:deform en}) after making the linear dependence of the stress from the strain explicit according to Hooke’s law (\ref{eq:Hooke}). Our elastic tensor is partially built from the metric, but it also constitutes an additional source together with the ordinary matter/energy term. \emph{In vacuo} the Bianchi identities bring the Einstein tensor to zero implying the conservation of ${T_e}_{\mu\nu}$, whereas in the presence of matter/energy the whole quantity $\kappa\,  T_{\mu\nu} +{T_e}_{\mu\nu}$ is conserved, signifying the possibility of a transfer of energy between the matter and the strain, which is a well-known subject in classical physics.

\subsection{A Friedmann-Lema\^{i}tre-Robertson-Walker universe}
\begin{figure}[!ht]
\begin{center}
\includegraphics[width=129 mm]{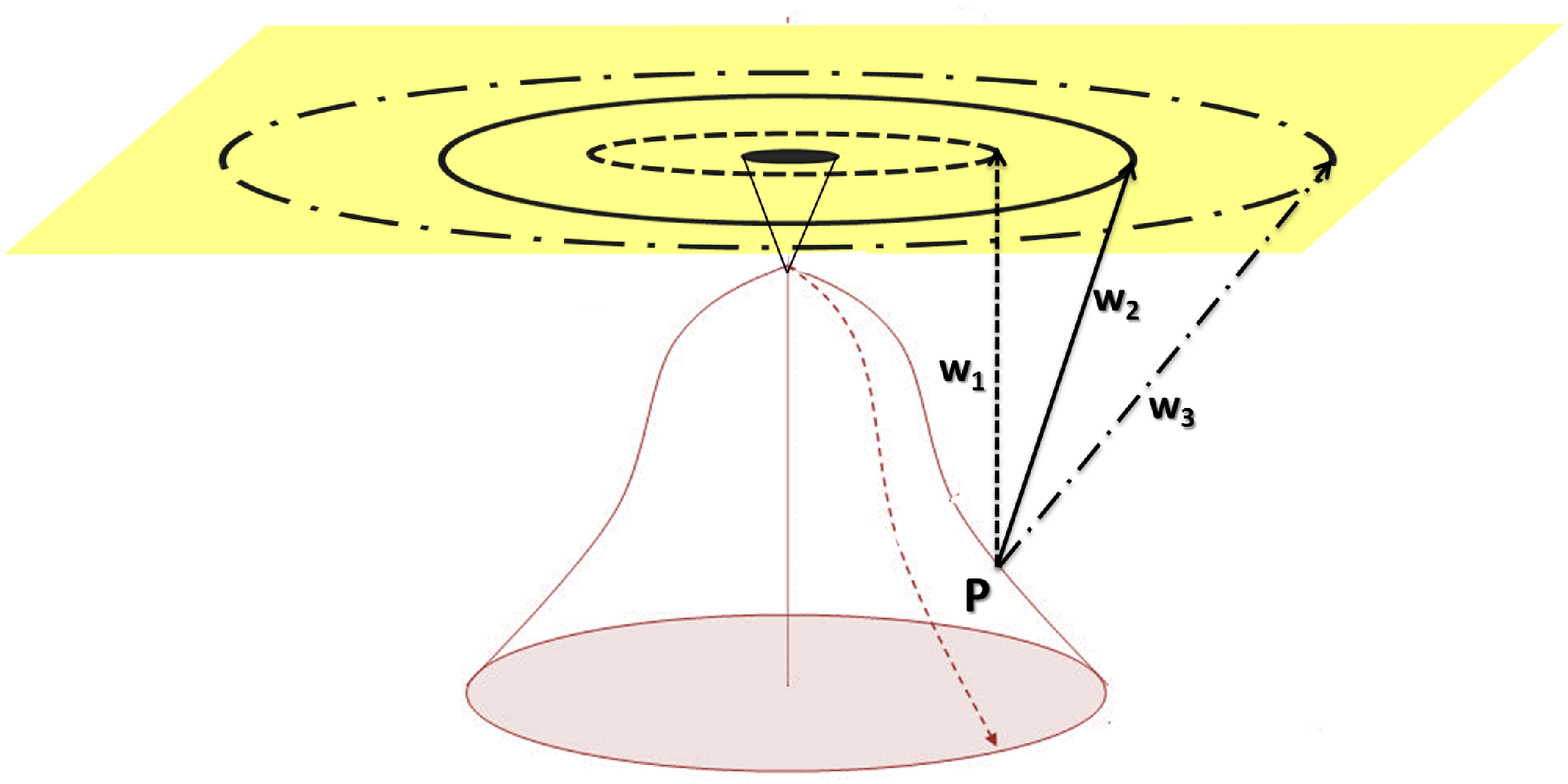}
\caption{A homogeneous and isotropic expanding universe, obtained by continuous deformation from a flat Euclidean manifold with a defect (the black disc on the reference manifold which maps to a single point on the natural manifold). The arrows correspond to different deformation strategies}
\label{fig:gauge}
\end{center}
\end{figure}

The strategy described so far has been applied to the study of the properties of the universe as a whole, assuming a global Robertson-Walker symmetry, i.e. homogeneity and isotropy in space (over the scale of hundreds of megaparsecs (Mpc)). The result is visually summarized in Fig.\ref{fig:gauge}, where the bell-shaped surface represents the space-time of our Friedmann-Lema\^{i}tre-Robertson-Walker (FLRW) universe, including a defect (`initial' singularity) and displaying (horizontal cross-sections) an expanding space, whose expansion rate is initially decreasing, then accelerating.

The flat surface on the top of the figure represents the Euclidean manifold from which, according to SSC, the real space-time has been obtained by deformation induced by the presence of a defect. The cosmic time is measured from the singularity (the defect) along the generators of the bell-shaped manifold (dashed line). The circles and the arrows shown in the figure manifest the peculiar gauge choices corresponding to different deformation strategies.

\section{The curvature-plus-strain method applied to \\ three-dimensional~solids}
We are now demonstrating that the method outlined in the previous pages might be successfully applied to three-dimensional material continua. Obviously, the time dimension is off the table: forgetting about space-time, we remain with a space in three dimensions with no Lorentz signature. However, the idea which we are trying to convey is that solids can be seen as three-dimensional manifolds that turn from their unstrained reference state to a curved natural configuration. We represent the unstrained state by means of a flat manifold, whose metric is determined hereafter. This is indeed a non-conventional interpretation of the theory of elasticity. The non-conventional aspect is not in the tensorial machinery, which was originally developed having continuum mechanics in mind, but in the use of the Einstein-Hilbert Lagrangian density with the addition of the deformation energy density. We may look at our approach as at an exercise, in which we shed light on the relations existing between two disparate branches of physics: cosmology and continua mechanics.

Hence, we are about to consider solid bodies with particularly symmetric shapes, as well as loaded symmetrically. Notably, we will develop the case of central symmetry, i.e. a spherical cavity immersed in an infinite medium, where the stresses originate from uniform and isotropic loadings which obey the symmetry of the body. For this case the classical theory of elasticity provides exact solutions that we are using to prove our thesis.

\subsection{Metric tensors for the natural and reference manifolds}
The natural manifold is a three-dimensional material continuum, which in general is not flat, but it has assumed a curvature as a consequence of some action applied to it. The reference manifold is an ordinary three-dimensional flat space: with no loss of generality we assume it coincides with $ \mathbb{R}^3$, described with a convenient set of coordinates, spherical in our case. Consequently, the embedding space has to be at least in four dimensions to contain the manifolds described above: the simplest possible is perhaps the four-dimensional Euclidean space $ \mathbb{R}^4$ .

In the fashion of the SSC, when a solid is unstrained its metric is flat, and in spherical polar coordinates this means:
\begin{equation}
\label{eq:spherical}
\mbox{d} s^2_{0}=\mbox{d} r^2+r^2\left( \mbox{d} \vartheta^2 +\sin^2 \vartheta\, \mbox{d} \varphi^2\right)
\end{equation}
Because the loading respects the central symmetry of the sphere, after deformation the symmetry is preserved, even though dimensions are slightly scaled. Fortunately, only one scale factor is needed to describe the metric, and it is function of the only coordinate $r$: we call it $f=f(r)$. The freedom in choosing the coordinate set offers many possibilities for the explicit form of the line element, always preserving the symmetry. Here we choose:
\begin{equation}
\label{eq:NATURAL}
\mbox{d} s^2=f(r)\,\mbox{d} r^2+r^2  \left( \mbox{d} \vartheta^2 + \sin^2 \vartheta \,\mbox{d} \varphi^2  \right)
\end{equation}

Consider now the reference manifold. It is flat, and its metric $E_{\mu\nu}$ is the tensor needed to represent the length of a line element before deformation. The $E_{\mu\nu}$ is indeed Euclidean: however, its explicit expression in terms of the coordinates used on the natural manifold involves knowing how the material was deformed.

Fig.\ref{fig:gauge} gives a clear idea of the problem. Point $P$ on the natural manifold may correspond to infinitely many points on the reference manifold. For example, assuming that $P$ lay on the solid circle before any stress was applied, the material shrank more than if it were on the dashed circle, but less than if it were on the dash-dot one. The various ways to go from the strained solid backward to the reference flat state depend on the strategy of deformation adopted. This strategy is represented by the function $w=w(r)$, which in practice establishes a one-to-one correspondence between points on the two manifolds. The choice for the correspondence is in principle arbitrary, but it can be determined by looking for the less energy consuming strategy.
The coloured arrows in Fig.\ref{fig:gauge} are the effective vectors of a displacement field $\vec{u}$: we may think of them as strings that pull points of the flat manifold, moving them to the natural configuration. Therefore, the same line element on the natural manifold may have several corresponding line elements on the reference manifold. Of course we must fix the form that $w$ assumes, eventually. In conclusion, the function $w$ simply defines the scale of the line element on the flat reference space, that has to be compared with the corresponding line element of the natural frame.
Summing up, the metric $E_{\mu\nu}$ of the reference manifold is
\begin{equation}
\label{eq:REFERENCE}
\mbox{d} s_r^2={w'}^2(r)\,\mbox{d} r^2+{w}^2(r)\left( \mbox{d} \vartheta^2 + \sin^2 \vartheta \,\mbox{d} \varphi^2  \right)
\end{equation}
where $'$ stands for the total derivative with respect to $r$.

\subsection{Lagrangian density}
The Lagrangian density $\mathcal{L}$ is indeed similar to (\ref{eq:lag dens SSC}), but with the more general $ \sqrt{|g|}$ instead of $ \sqrt{-g}$: $\mathcal{L}=\left(  R + W \right) \sqrt{|g|}$. The two ingredients $R$ and $W$ are ready to be built, because the metrics are already well defined.

The curvature of the reference manifold is obviously zero, whereas the curvature of the natural manifold is directly computed using the metric tensor $g_{\mu\nu}$ of (\ref{eq:NATURAL}):
\begin{equation}
\label{eq:SCALAR CURVATURE}
R=\frac{2rf^{\prime }+2f^{2}-2f}{r^{2}f^{2}}
\end{equation}

On the other hand, applying (\ref{eq:metrics}), the non-zero components of the strain tensor in covariant form are
\begin{eqnarray}
\varepsilon _{rr} &=&\frac{f-w^{\prime 2}}{2} \label{eq:epsilonrr1}\\
\varepsilon _{\vartheta \vartheta } &=&\frac{r^{2}-w^{2}}{2} \label{eq:epsilontt1}\\
\varepsilon _{\varphi \varphi } &=&\frac{r^{2}-w^{2}}{2}\sin ^{2}\vartheta
\end{eqnarray}
To raise one subscript of the strain tensor in order to obtain the mixed form  $ \varepsilon_{\mu}^{ \nu}$, let us use the inverse metric $ g^{\mu \nu}$:
\begin{eqnarray}
\varepsilon _{r}^{r} &=&\frac{f-w^{\prime 2}}{2f} \\
\varepsilon _{\vartheta }^{\vartheta } &=&\frac{r^{2}-w^{2}}{2r^{2}}=\varepsilon
_{\varphi }^{\varphi }
\end{eqnarray}
In this way, the scalar strain is immediately built:
\begin{equation}
\varepsilon =\varepsilon_\alpha^\alpha=\frac{f-w^{\prime 2}}{2f}+\frac{r^{2}-w^{2}}{r^{2}}
\end{equation}
Applying again the inverse metric $ g^{\mu \nu}$ to $ \varepsilon_{\mu}^{ \nu}$, we easily get $ \varepsilon^{\mu \nu}$. So, we can compute the second degree scalar
\begin{equation}
\varepsilon _{\alpha \beta }\varepsilon ^{\alpha \beta }=\frac{\left(
f-w^{\prime 2}\right) ^{2}}{4f^{2}}+\frac{\left( r^{2}-w^{2}\right) ^{2}}{%
2r^{4}}
\end{equation}
The deformation energy is then obtained by casting these results into (\ref{eq:deform en}). Eventually, the determinant of the metric yields
\begin{equation}
\sqrt{|g|}=\sqrt{f}\,r^2\,\sin\vartheta
\end{equation}
And the Lagrangian density is:
\begin{multline}\label{eq:LAG DENS}
\frac{\mathcal{L}}{\sin \vartheta }=\Biggl[ \frac{2rf^{\prime }+2f^{2}-2f}{r^{2}f^{2}}+%
\frac{1}{2}\lambda \left( \frac{f-w^{\prime 2}}{2f}+\frac{r^{2}-w^{2}}{r^{2}}%
\right) ^{2}+\\+\mu \left( \frac{\left( f-w^{\prime 2}\right) ^{2}}{4f^{2}}+%
\frac{\left( r^{2}-w^{2}\right) ^{2}}{2r^{4}}\right) \Biggr]
 \sqrt{f}r^{2}
\end{multline}

\section{Results: extremization of the Lagrangian}
From (\ref{eq:LAG DENS}), we could develop the so-called Lagrange equations. Nonetheless, these equations would be highly non-linear and very difficult to solve. There exists a simpler way to verify that a Lagrangian density like (\ref{eq:lag dens SSC}) accounts also for the deformation of a three-dimensional solid. Recall that, in the case of a field theory, Hamilton's principle can be stated as in Landau and Lifshitz \cite{landau field} (pp.287-288):
\begin{equation}
\label{eq:HAMILTON}
\delta S= \delta\left( \int_{\Omega^{(3)}} \mathcal{L} \,\mbox{d}^3x \right)=0
\end{equation}
Hamilton's principle (\ref{eq:HAMILTON}) states that the action must be extremal (a minimum in general, but not always so): then, it also involves that in order for $\delta S$ to be zero,  the Lagrangian density $\mathcal{L}$ is necessarily extremal. This is why Hamilton's principle will be restated as
\begin{equation}
\label{eq:HAMILTON2}
\delta \mathcal{L} = 0
\end{equation}
which means that when we perform the variation of the Lagrangian density we find zero.

Where $w$ and $f$ are being derived from? We can use the exact solutions worked out within the classical elasticity theory. In particular, we wish to determine the strain tensor for a hollow spherical cavity of radius $b$ subjected to a spherically symmetric loading, say a uniform internal pressure $p$. The walls of the cavity are assumed to be infinitely thick, so that no stress acts from outside.
We use the spherical polar coordinates $(r,\vartheta,\varphi)$, and note that the displacement vector $\vec{u}$ is function of the only variable $r$. This implies $\vec{\nabla} \times \vec{u}=0$, and the Navier-Cauchy equilibrium equation yields: $\vec{\nabla} \cdot \vec{u}=\mbox{constant}$. We integrate this equation, and use the strain-displacement and stress-strain relations (following the approach outlined in Landau and Lifshitz \cite{landau el}, pages 20-21). Imposing boundary conditions as follows
\begin{equation}
\label{eq:boundary sphere}
\sigma_r^r=\left\{
\begin{array}{rl}
0 & \mbox{if } r \rightarrow \infty \\
p & \mbox{if } r=b
\end{array}
\right.
\end{equation}
we find that the strain tensor is
\begin{eqnarray}
\varepsilon^r_r=\frac{pb^3}{2\mu}\frac{1}{r^3} \label{eq:strain_mix1}\\
\varepsilon^{\vartheta}_{\vartheta}=\varepsilon^{\varphi}_{\varphi}=-\frac{pb^3}{4\mu}\frac{1}{r^3} \label{eq:strain_mix2}
\end{eqnarray}
Everything here is expressed using coordinates defined with respect to the unstrained reference frame. This is the common practice when studying elasticity; in the cosmological application instead, the natural choice is to use the coordinates defined for the actual universe where we live in, which correspond to the strained rather than the reference frame. In any case, as we saw before, for the final purpose the two choices are perfectly equivalent.
Comparing (\ref{eq:strain_mix1}) and (\ref{eq:strain_mix2}) in covariant form with (\ref{eq:epsilonrr1}) and (\ref{eq:epsilontt1}) we get
\begin{eqnarray}
w=\sqrt{\frac{2r^{3}\mu +pb^{3}}{2\mu r}} \\
f=\frac{8pb^{3}r^{3}\mu +9p^{2}b^{6}+16r^{6}\mu ^{2}}{8\mu
r^{3}(2r^{3}\mu +pb^{3})}
\end{eqnarray}
In practice, we have rewritten the exact solutions in terms of $f$ and $w$, which we now cast into (\ref{eq:LAG DENS}), thus finding an explicit expression for the Lagrangian density. Let us call it $\overline{\mathcal{L}}$.
Performing the variation on $\overline{ \mathcal{L} }(r)$, we should find that $\delta \overline{ \mathcal{L} } = 0$ holds up to higher order terms. This fact is apparent if we consider that the SSC is not linear, but the values for $f$ and $w$ we cast inside are obtained from a linear theory.

\begin{figure}[!ht]
\centering
\includegraphics[width=129 mm]{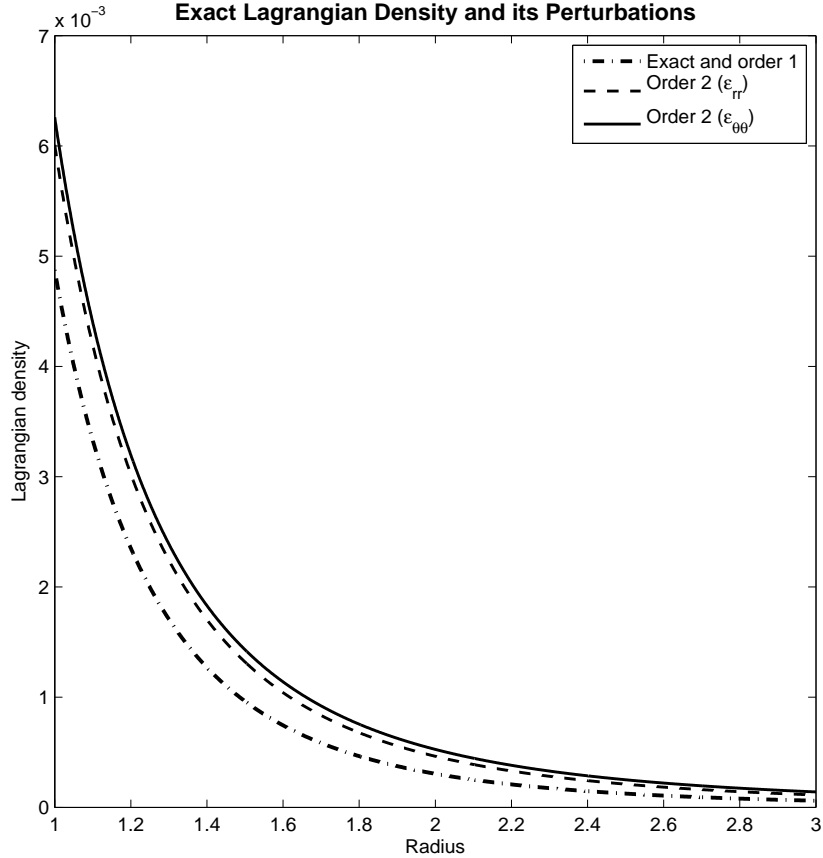}
\caption{Lagrangian density as a function of the radius (dash-dot line) and perturbations thereof expanded up to second order (the dashed curve represents the case where $\varepsilon_{rr}$ is perturbed, and the solid one the case where $\varepsilon_{\vartheta\vartheta}$ is perturbed). When we expand the perturbed Lagrangians only up to first order, the curves coincide with the dash-dot one}
\label{fig:L}
\end{figure}

In other words, what we are proving now is that these solutions minimize the Lagrangian density. To do so, we vary the values of $f$ and $w$ by perturbing the exact solutions (\ref{eq:strain_mix1}) and (\ref{eq:strain_mix2}), and find that the Lagrangian density immediately increases. For instance, consider the  perturbation $\varepsilon_{rr}+\alpha(r)$, where $\alpha(r)$ is a small function (compared to $\varepsilon_{rr}$), and $\varepsilon_{\vartheta \vartheta}$ is left unperturbed. Proceeding as before we have
\begin{equation*}
\frac{r^{2}-w^{2}}{2} =-\frac{pb^3}{4\mu}\frac{1}{r} \quad \mbox{and} \quad \frac{f-w^{\prime 2}}{2} =\frac{pb^3}{2\mu}\frac{1}{r^3}+\alpha(r)
\end{equation*}
Solving $f$ and $w$, and plugging these values into $\overline{\mathcal{L}}(r)$ (expanding with respect to powers of $\alpha$) we find the perturbed Lagrangian $\tilde{\mathcal{L}}$:
\begin{equation*}
\tilde{\mathcal{L}}=\overline{ \mathcal{L} }+\alpha \cdot \left( 0\right)+\alpha^2 \cdot \left( \ldots\right) + \ldots
\end{equation*}
Although up to first order $\tilde{\mathcal{L}}=\overline{ \mathcal{L} }$, if we add a further term in the expansion, i.e. second order terms ($\alpha^2$ collects these terms, whereas $\alpha$ collects none), the Lagrangian does get bigger, indeed. These results are shown in Fig.\ref{fig:L}, where the computation has been repeated moving the perturbation onto the other non-zero strain component, i.e. $\varepsilon_{\vartheta \vartheta}+\alpha(r)$.

Our simple example with three-dimensional spherical symmetry corresponds in cosmology to a closed ever expanding FLRW universe. If we had treated a different symmetry, also the four-dimensional correspondence would have been to other cosmological models with different topologies of the initial singularity.

\section{Conclusions}
We have shown how a generalization of General Relativity, obtained exploiting the analogy with the ordinary linear theory of elasticity in three dimensions, is able to account for the global behaviour of the universe, and in particular to reproduce the accelerated expansion discovered in 1998 and 1999 observing the luminosity of type Ia supernovae (Riess et al. \cite{ries}, and Perlmutter et al. \cite{perl}). The situation is visually  described in Fig.\ref{fig:gauge} where successive spacelike cross-sections of space-time increase at a growing rate with the distance from the initial singularity (i.e. with cosmic time). This result is contained in ref.s \cite{radicella} and \cite{rst1}. 

Our result is obtained adding to the classical Einstein-Hilbert action for empty space-time a term accounting for the equivalent of a deformation energy, built on the strain induced in space-time by either the presence of texture defects (as some times happens in ordinary three-dimensional continua) or of matter/energy distributions.
The additional term lends itself also to a simple interpretation of the so called `dark energy' of cosmology. Working out formally the equation of state of the strain term as if it were a fluid in the FLRW universe, we see that the $w$ parameter is negative (for big enough values of the scale parameter $a$), which corresponds to a negative pressure (ref.\cite{radicella}, sec.5): a property which is ascribed to the dark energy. \textbf{More precisely} we should write that the strain term resembles the `quintessence' of cosmology, since the parameter of state is not equal to -1 (cosmological constant) but corresponds to a function of the scale factor. The fact that we refer to a strain fluid rather than to a solid is not important here, because the solutions we find in four dimensions are the equivalent of equilibrium configurations so that the distinction is only in the explicit form of the energy density. The difference between the classical picture of an evolution in time and the four-dimensional manifold of GR is in that an evolution scenario would require at least a fifth dimension with an evolution parameter replacing the role of the independent time of classical dynamics. This is why we say that any global solution for the configuration of space-time is an `equilibrium' because it does not depend on any further evolution parameter, just as it happens for the equilibria of three-dimensional classical systems.
The strained state is represented by the strain tensor, intended, as it is the case in the theory of elasticity, as half the difference between the intrinsic metric tensor of the actual deformed continuum and the metric tensor of the initial unaffected manifold. Whenever the strain is identically zero we have a flat manifold, i.e. a continuum whose intrinsic geometry is Euclidean. In the case of space-time, gravitational effects are included in the strain tensor, i.e. in the non-trivial part of the metric tensor. Once the action integral has been generalized in the way outlined above, the rest is deduced applying Hamilton's principle.
Next, we have shown that the same approach adopted at the cosmological scale is viable also in three dimensions to treat ordinary problems of elasticity. Here too, even though more direct approaches are often at hands, the equilibrium configurations of elastic continua may be deduced from the three-dimensional version of the SSC action integral used in cosmology. The reference to cosmology is rather appealing, but we would like to stress that the method we have outlined is indeed an extension of GR and can be applied to typical problems of gravitation with different symmetries. It has for instance been applied recently to cylindrical symmetry in four dimensions, i.e. to the Schwarzschild problem (spherical symmetry in space) \cite{rst2} and will be applied to screw symmetry (rotating systems). We hope that the corresponding lower dimensional symmetries can help in solving the GR problems.

Summing up, we have seen an example of how the universal nature of physical laws is manifested establishing a connection between the universe as a whole, or at least space-time that is the arena on which the cosmic drama is played, and the everyday familiar description of the properties of deformable solids.

\end{document}